\documentclass[iop]{emulateapj}

\def\M1{M_1^\prime}
\def\Msun{M_\odot}

\gdef\Vdisp{$\sigma$}

\gdef\Mstar{M$_{\rm star}$}
\gdef\Mden{$\Sigma$}
\gdef\Re{$R_{e}$}

\slugcomment{Submitted to ApJ Letters}

\shorttitle{}
\shortauthors{Wake et al.}

\begin{document}

%% LaTeX will automatically break titles if they run longer than
%% one line. However, you may use \\ to force a line break if
%% you desire.

%%\title{Which is the best indicator of a galaxy's color: stellar mass, velocity dispersion, surface mass density or morphology?}
\title{Revealing velocity dispersion as the best indicator of a galaxy's color, compared to stellar mass, surface mass density or morphology}

%% Use \author, \affil, and the \and command to format
%% author and affiliation information.
%% Note that \email has replaced the old \authoremail command
%% from AASTeX v4.0. You can use \email to mark an email address
%% anywhere in the paper, not just in the front matter.
%% As in the title, use \\ to force line breaks.

\author{David. A. Wake\altaffilmark{1},
Pieter G.\ van Dokkum\altaffilmark{1}
Marijn Franx\altaffilmark{2},
}

\altaffiltext{1}{Department of Astronomy, Yale University, New Haven, CT 06520}
\altaffiltext{2}{Sterrewacht Leiden, Leiden University, NL-2300 RA Leiden,
The Netherlands.}

%% Mark off your abstract in the ``abstract'' environment. In the manuscript
%% style, abstract will output a Received/Accepted line after the
%% title and affiliation information. No date will appear since the author
%% does not have this information. The dates will be filled in by the
%% editorial office after submission.

\begin{abstract}

Using data of nearby galaxies from the
Sloan Digital Sky Survey we investigate whether stellar mass (\Mstar),
central velocity dispersion (\Vdisp), surface mass density (\Mden), or
the Sersic $n$ parameter is
best correlated with a galaxy's rest-frame color. Specifically,
we determine how
the mean color of galaxies varies with one parameter when another is
fixed.  When \Mstar\ is
fixed we see that strong trends remain with all other parameters,
whereas
residual trends are weaker when \Mden, $n$, or \Vdisp\ are fixed.
Overall \Vdisp\ is the best
indicator of a galaxy's typical color, showing the largest residual color dependence when any of the other three parameters are fixed, and \Mstar\
is the poorest. 
Other studies have indicated that both the halo and black hole properties
are better correlated with \Vdisp~than with \Mstar, \Mden~or $n$.
Therefore, our
results are consistent with a picture where a galaxy's star formation
history and present star formation rate are determined to some
significant degree by the current properties and assembly history of
its dark matter halo and/or the feedback from its central super
massive black hole. 

\end{abstract}

%% Keywords should appear after the \end{abstract} command. The uncommented
%% example has been keyed in ApJ style. See the instructions to authors
%% for the journal to which you are submitting your paper to determine
%% what keyword punctuation is appropriate.


%% From the front matter, we move on to the body of the paper.
%% In the first two sections, notice the use of the natbib \citep
%% and \citet commands to identify citations.  The citations are
%% tied to the reference list via symbolic KEYs. The KEY corresponds
%% to the KEY in the \bibitem in the reference list below. We have
%% chosen the first three characters of the first author's name plus
%% the last two numeral of the year of publication as our KEY for
%% each reference.

%% Authors who wish to have the most important objects in their paper
%% linked in the electronic edition to a data center may do so by tagging
%% their objects with \objectname{} or \object{}.  Each macro takes the
%% object name as its required argument. The optional, square-bracket 
%% argument should be used in cases where the data center identification
%% differs from what is to be printed in the paper.  The text appearing 
%% in curly braces is what will appear in print in the published paper. 
%% If the object name is recognized by the data centers, it will be linked
%% in the electronic edition to the object data available at the data centers  
%%
%% Note that for sources with brackets in their names, e.g. [WEG2004] 14h-090,
%% the brackets must be escaped with backslashes when used in the first
%% square-bracket argument, for instance, \object[\[WEG2004\] 14h-090]{90}).
%%  Otherwise, LaTeX will issue an error. 

\section{Introduction}

It is well established that the stellar populations of galaxies in the
nearby Universe correlate with their luminosities and masses, such that
the stars in
more massive galaxies are on average older and more metal-rich \citep[e.g.,][]{Bower92,Blanton03, Kauffmann03}. The old ages
of the most luminous galaxies are somewhat puzzling, as it implies that
there must be some
mechanism responsible for shutting off their star formation \citep[e.g.,][]{Kauffmann00,Croton06,Naab07,Keres05}.
Another puzzle is that 
the correlation between star formation history and mass is complex: low
mass galaxies are typically younger and high mass galaxies are typically
older, with a bimodal transition region at $\sim 3\times 10^{10}\,\Msun$
\citep{Kauffmann03}. This bimodality and the associated transition
mass scale have been the subject of intense debate in the past years
\citep[e.g.,][]{bundy:06,Peng10,brammer:11}.

An intriguing possibility is that luminosity and mass are not the
``right'' parameters to interpret galaxy evolution, and that there
exist other parameters that show more straightforward correlations with
stellar population parameters. As demonstrated by \citet{Kauffmann03,Kauffmann06}, the star formation histories of galaxies show less scatter when the structure
of galaxies is taken into account. In particular, the ages and star formation rates of galaxies
are better correlated with their
surface densities \Mden\ (which is proportional to
$M_*/r_e^2$ with $r_e$ the size of the galaxy) than with mass alone.
\citet{Franx08} find that the strong correlation between color
and $M_*/r_e^2$ (and $M_*/r_e$) persists all the way to $z\sim 2$.
Similar trends have been found for velocity dispersion
and for the \citet{Sersic} index $n$ \citep[e.g.,][]{Blanton03, Bell08, Wuyts11,Dokkum11,Bell11}.

Although it seems clear that the correlation between color and mass
is weaker than the correlations between color and \Mden, \Vdisp, or
$n$, it is not clear which of these parameters is the best predictor
of a galaxy's color. This is important to establish as it
provides information on the physical
processes that govern galaxy evolution. As an example, if $n$ best
correlates with color \citep[as suggested by][]{Bell11}
it may imply that the merger history determines
the star formation history, whereas if \Vdisp\ is the key parameter
it would suggest that the star formation history is
influenced (or even determined) by the
dark matter halo or the central supermassive black hole.

In this Letter we investigate which of the four parameters \Mstar,
\Mden, \Vdisp, or $n$ is most closely correlated
with the star formation history, as parameterized by the color. 
We determine this by fixing each parameter in turn and measuring to
what extent the color depends on the other three parameters. The
homogeneous, large datasets required for this study are now available
from the 7th data release of the Sloan Digital Sky Survey \citep[SDSS DR7;][]
{Abazajian09}.

\section{Data}

The galaxy data used in this analysis are gathered from the seventh data release of the SDSS \citep{Abazajian09}. We begin with the Large Scale Structure samples of the DR7 NYU Value Added Galaxy catalogue \citep[VAGC][]{Blanton05}. The sample we use has an $r$ band magnitude range of $14.5 < r < 17.6$. In addition the NYU VAGC gives k-corrected (to z=0.1) absolute magnitudes \citep{Blanton03}, velocity dispersion measurements from the Princeton Spectroscopic pipeline, and circularized sersic fits for each galaxy all of which we make use of in this analysis.

For estimates of the stellar mass we make use of the MPA/JHU DR7 value added catalog which provides stellar mass estimates based on stellar population fits to the SDSS photometry \citep{Kauffmann03, Salim07}. The overlap between the MPA/JHU and NYU VAGs is close to but to quite 100\% and so we remove the regions where they do not match from the analysis. We also remove any region of the survey that has a spectroscopic completeness less than 70\%. This leaves an area of 7640 deg$^2$ and a total sample of 521,313 galaxies.

The SDSS velocity dispersions are measured within the 3" diameter SDSS fiber. We correct to a common aperture of one eighth of an effective radius ($r_e$), the central velocity dispersion, using the relation $\sigma_0 = \sigma_{ap}(8r_{ap}/r_e)^{0.066}$ where $r_{ap}$ = 1."5 \citep{Cappellari06}. $r_e$ is taken from the best fitting circularized Sersic profile fit.

Throughout we use $u-r$ color from the $K$-corrected NYU VAGC absolute magnitudes. As already stated these are corrected to $z=0.1$; although
they are not quite $u-r$ at rest we will refer to them as
$u-r$ colors.
We also make use of the morphological classifications from Galaxy Zoo \citep{Lintott11} which provides multiple visual classifications for each galaxy in the SDSS spectroscopic sample. The parameter we use is the probability that a galaxy is an edge on disk ($P_{edge}$) \citep[see ][ for details]{Lintott11}.

\section{Dependence of color on \Mstar, \Vdisp, \Mden, and $n$}

\begin{figure}

\vspace{10.0cm}
\includegraphics{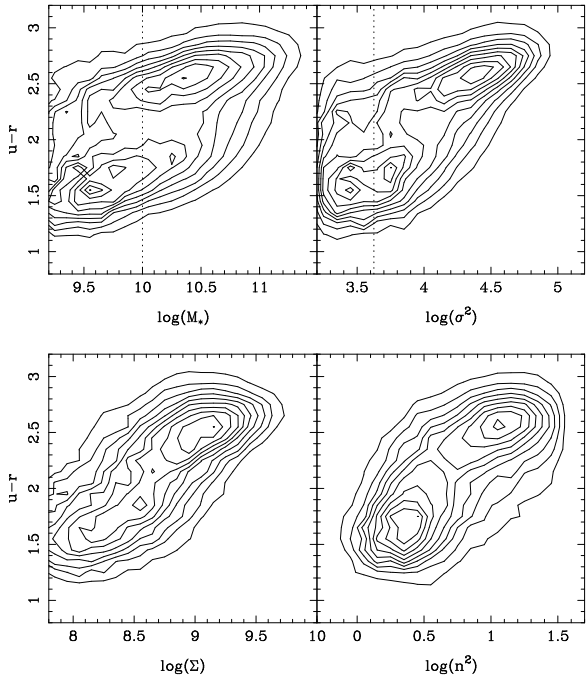}
\caption{\small
The $u-r$ color distribution as a function of \Mstar, \Vdisp$^2$,
\Mden~and $n^2$, for SDSS DR7 galaxies with $0.04<z<0.113$. The $u-r$ color
is positively correlated with all four parameters. The correlation with
\Mstar\ has large scatter, particularly near \Mstar$\sim3\times
10^{10}\,\Msun$. The dotted lines show the cuts we apply in \Mstar\ and \Vdisp\ to define our primary sample.
\label{fig:ColNsVdMdSM}}
\end{figure}

\begin{figure*}
\vspace{11.0cm}
\includegraphics{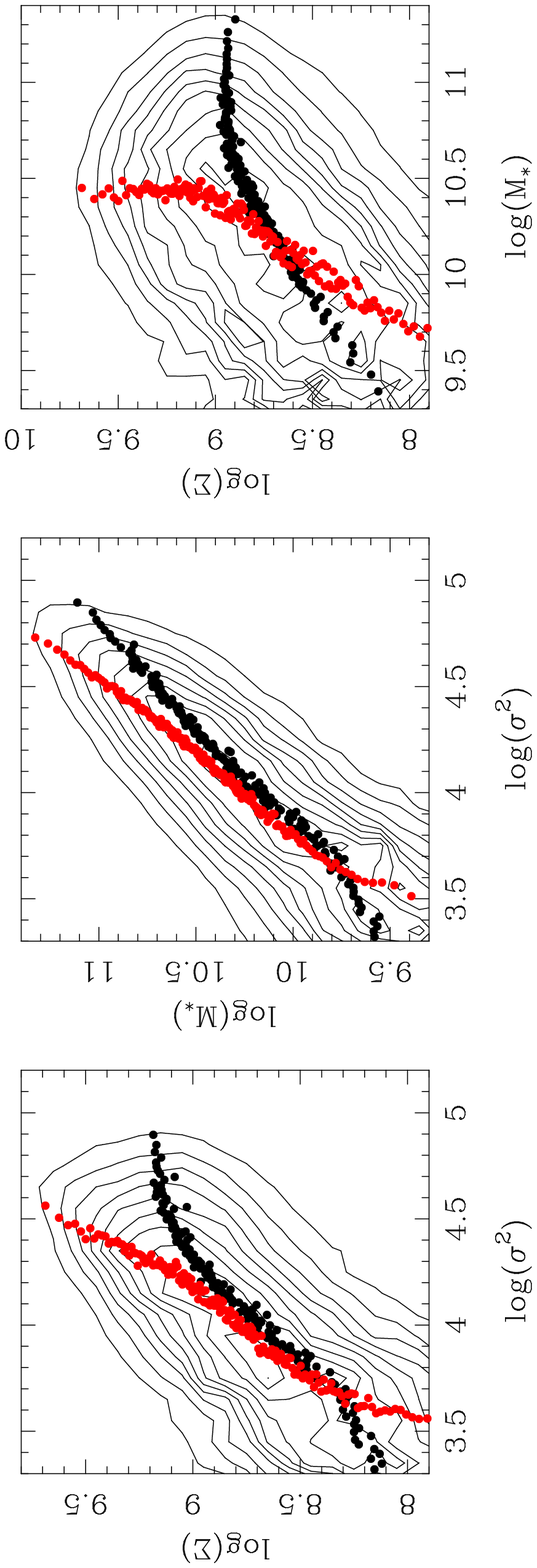}
\includegraphics{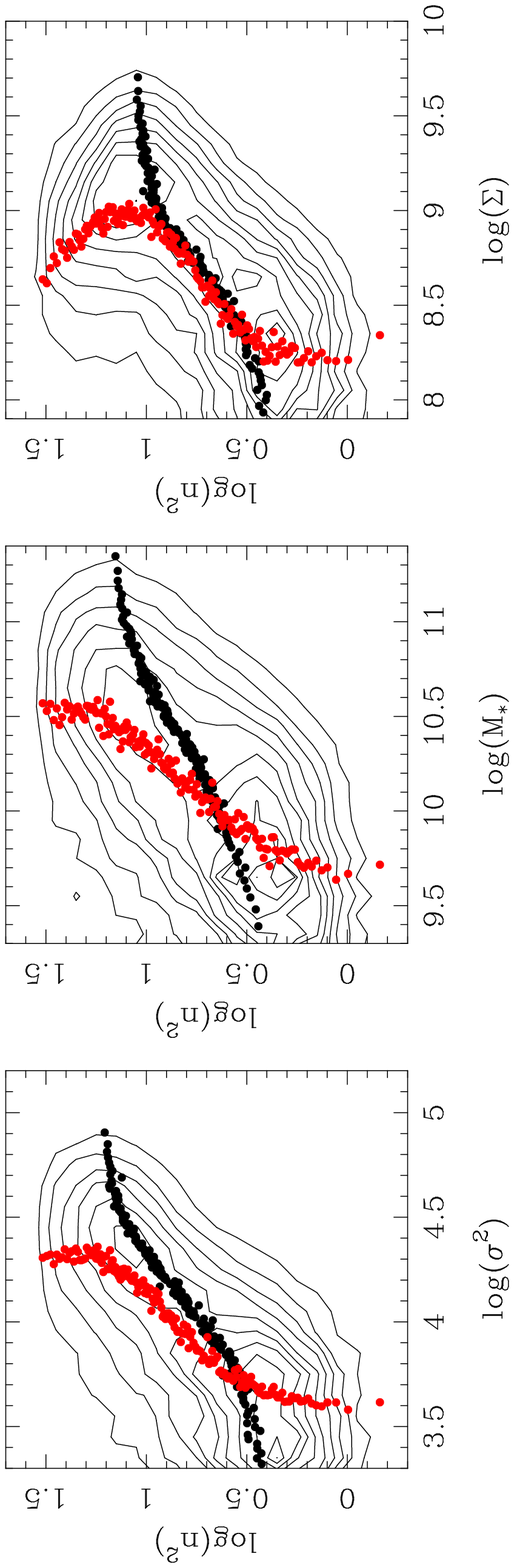}
\caption{\small The relationships between \Mstar, \Mden, \Vdisp$^2$ and $n^2$ for all SDSS DR7 galaxies with 0.02 $< z <$ 0.11. Galaxies are V/Vmax weighted to correct for the redshift dependent stellar mass completeness limit. The black points show the mean of y in bins of x containing 500 galaxies, where as the red points show the mean of x in bins of y. We can see that over a significant range of each parameter they are all approximately linearly proportional to one another.  
\label{fig:VdispMdenSM}}
\end{figure*}

\begin{figure*}

\vspace{21.0cm}
\includegraphics{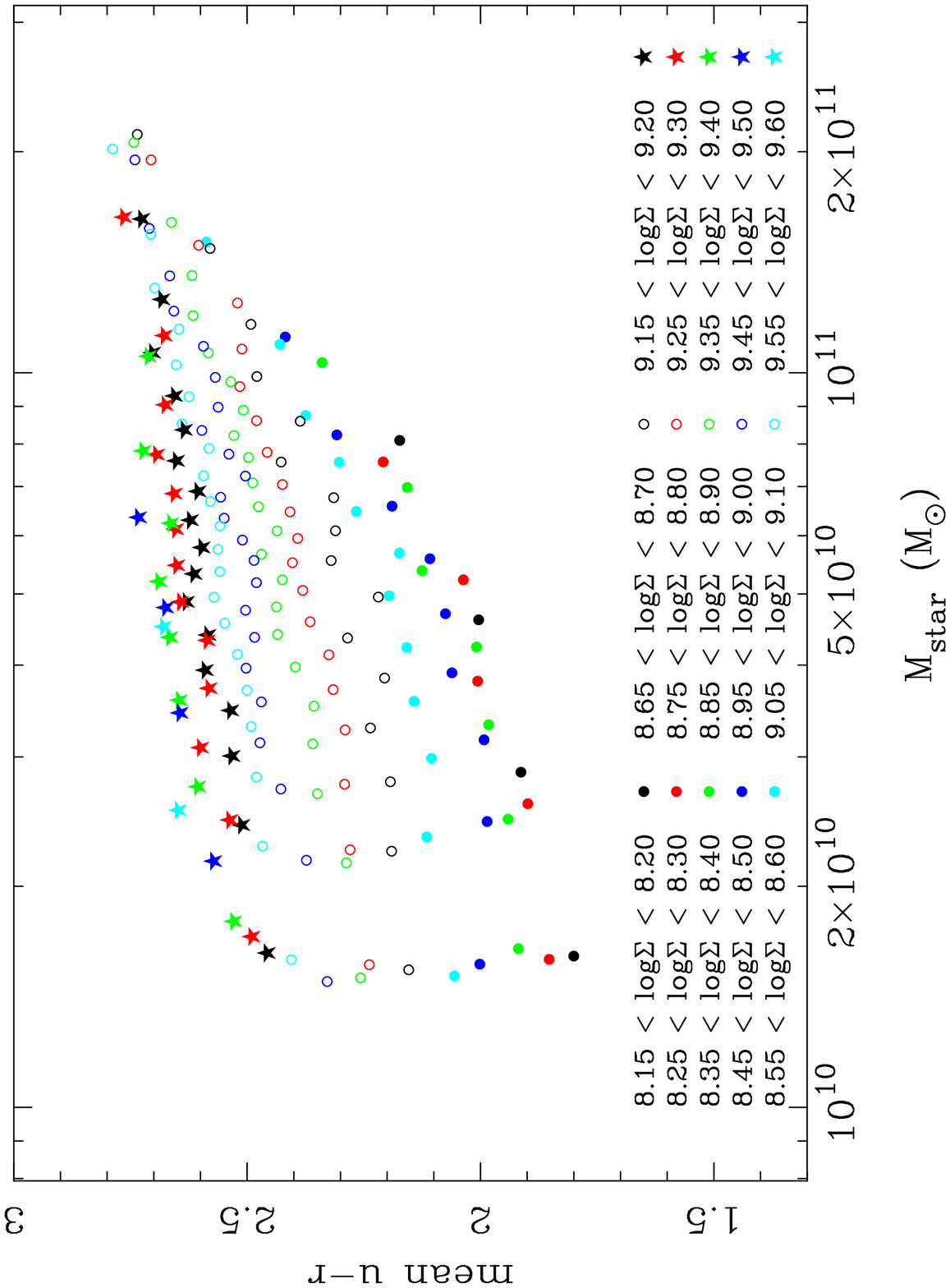}
\includegraphics{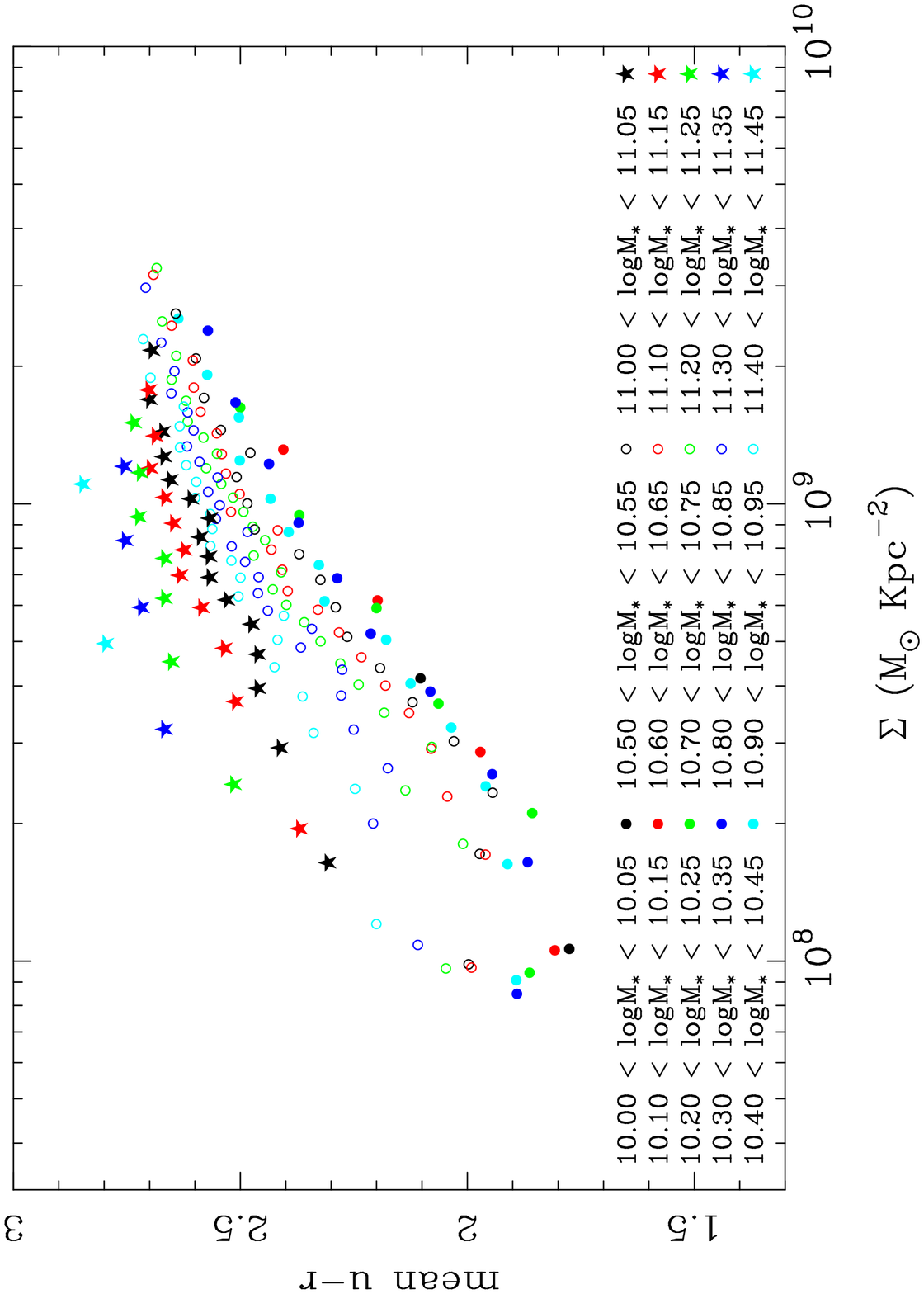}
\includegraphics{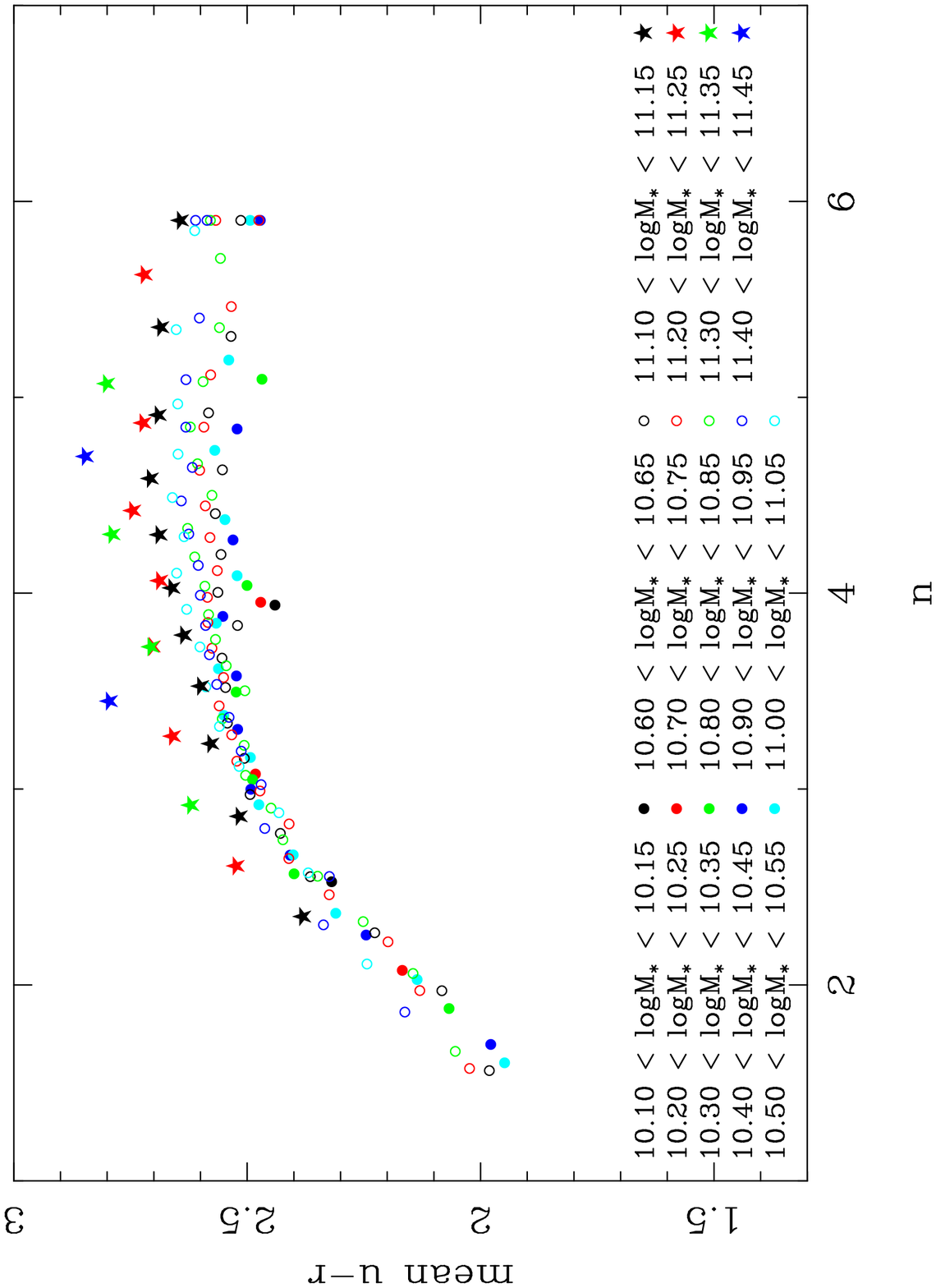}
\includegraphics{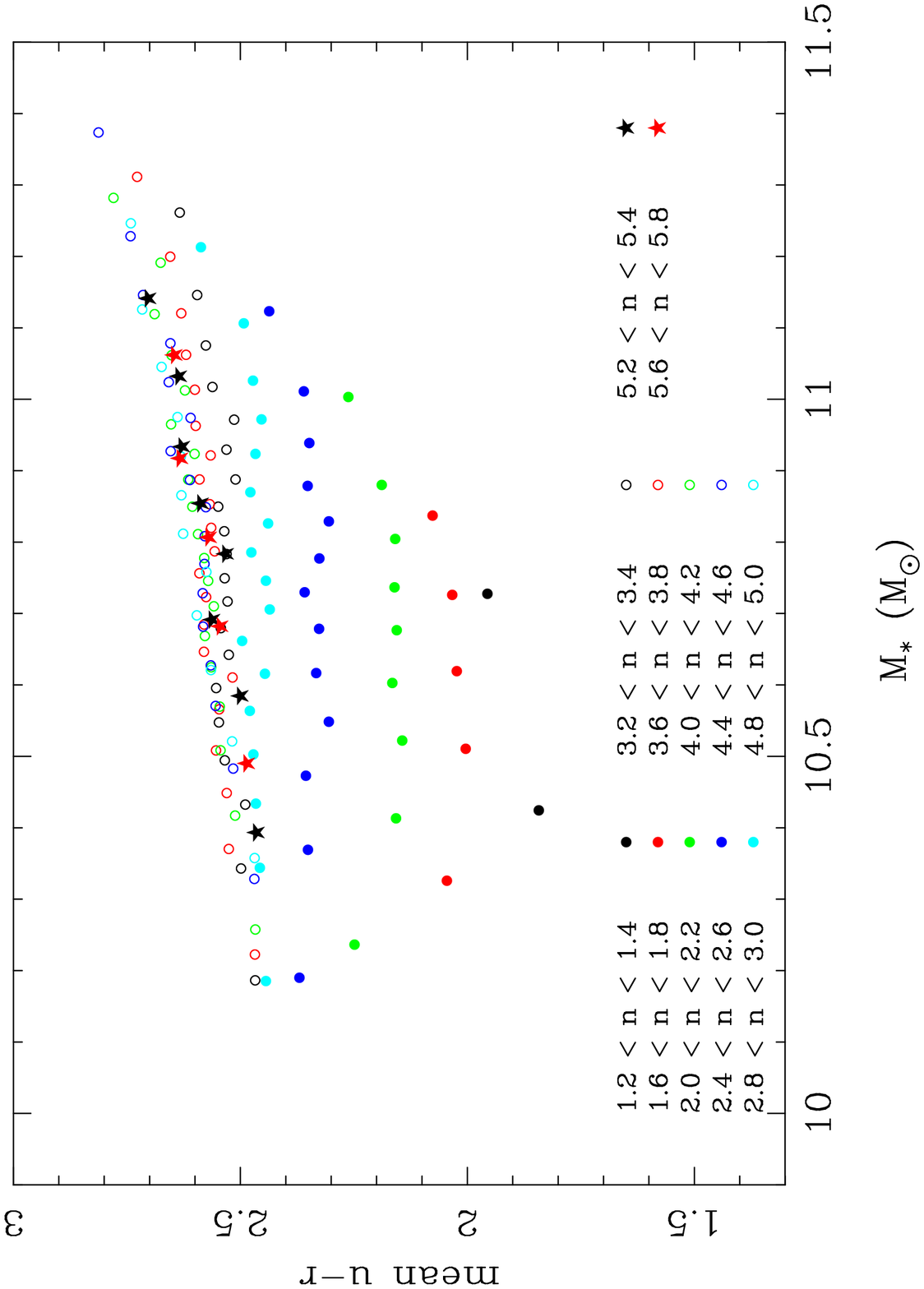}
\includegraphics{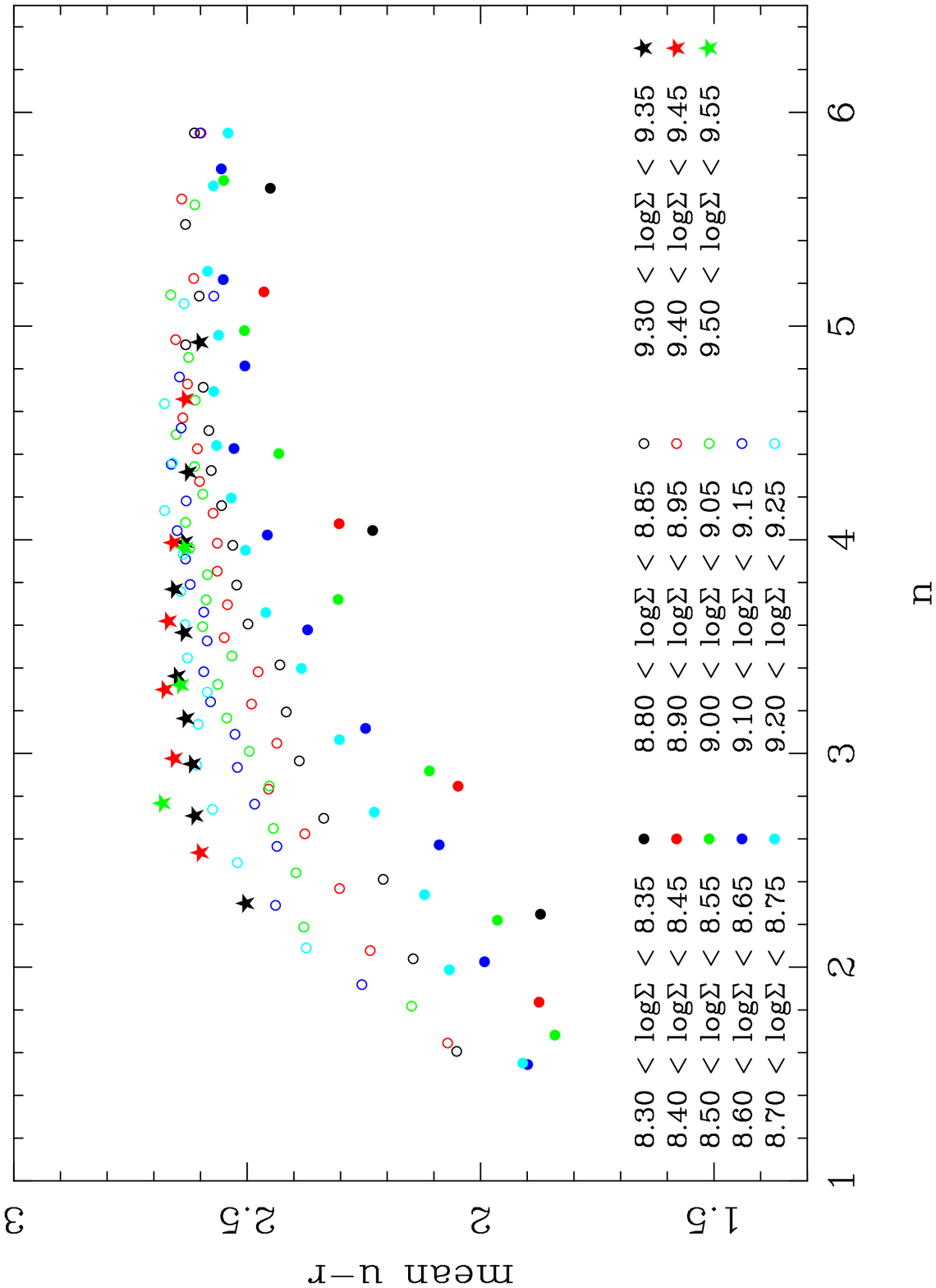}
\includegraphics{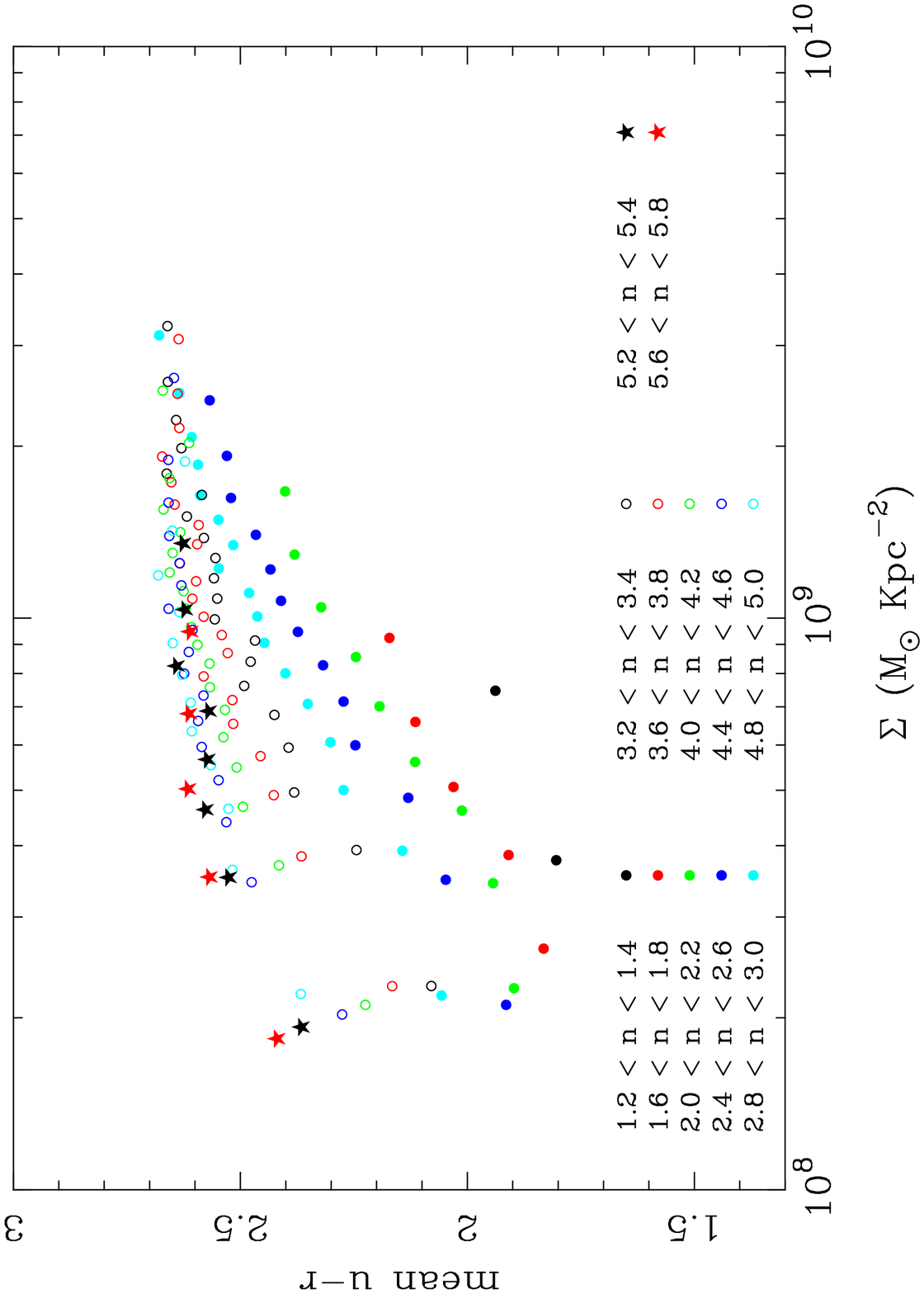}
\caption{\small The relationship between mean $u-r$ color and mass surface density at fixed stellar mass (top left), stellar mass at fixed mass surface density (top right), stellar mass at fixed sersic n (middle left), sersic n at fixed stellar mass (middle right), mass surface density at fixed sersic n (bottom left) and sersic n at fixed mass surface density (bottom right). 
\label{fig:colNsMdenSM}}
\end{figure*}

\begin{figure*}
\vspace{21.0cm}
\includegraphics{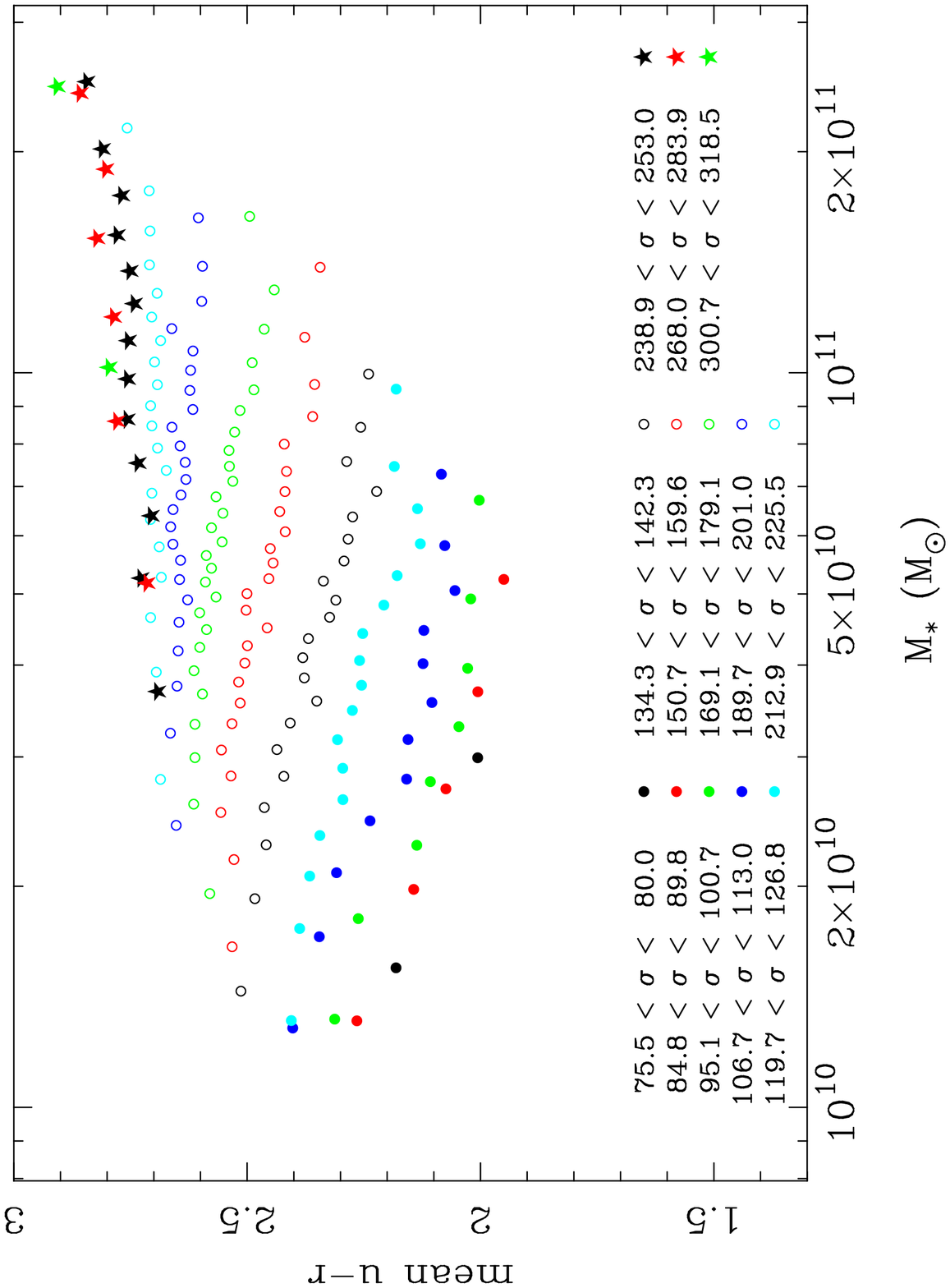}
\includegraphics{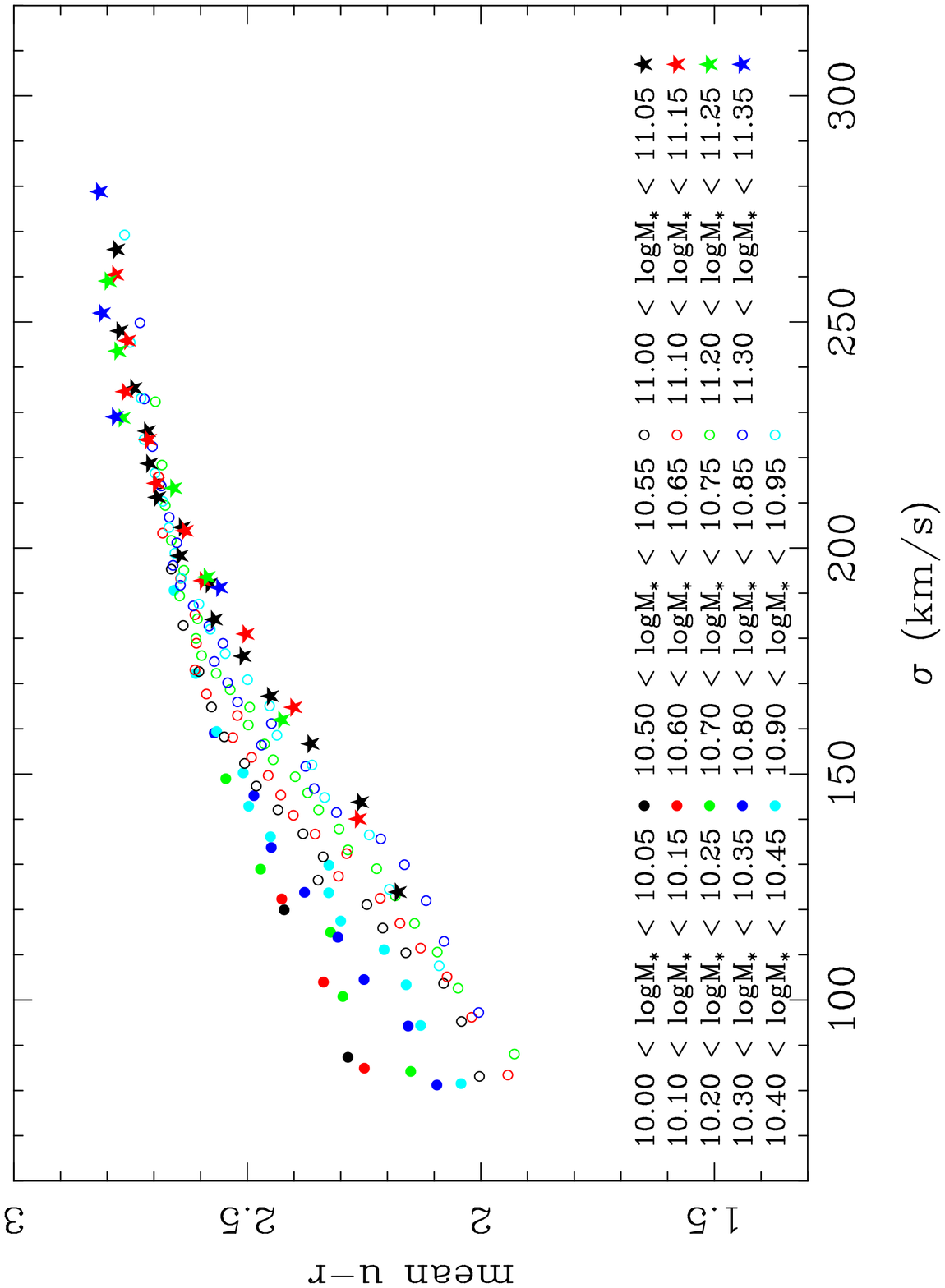}
\includegraphics{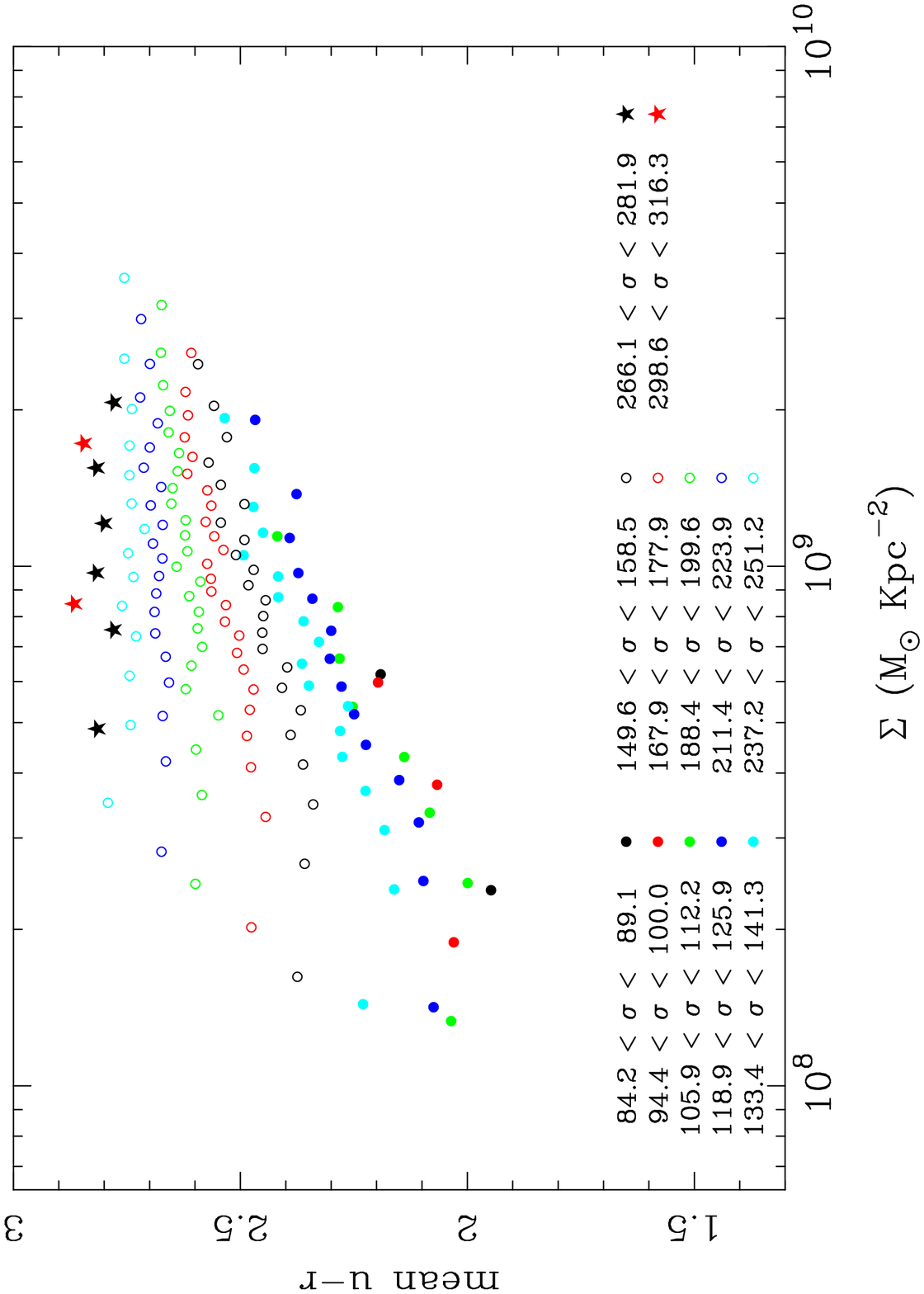}
\includegraphics{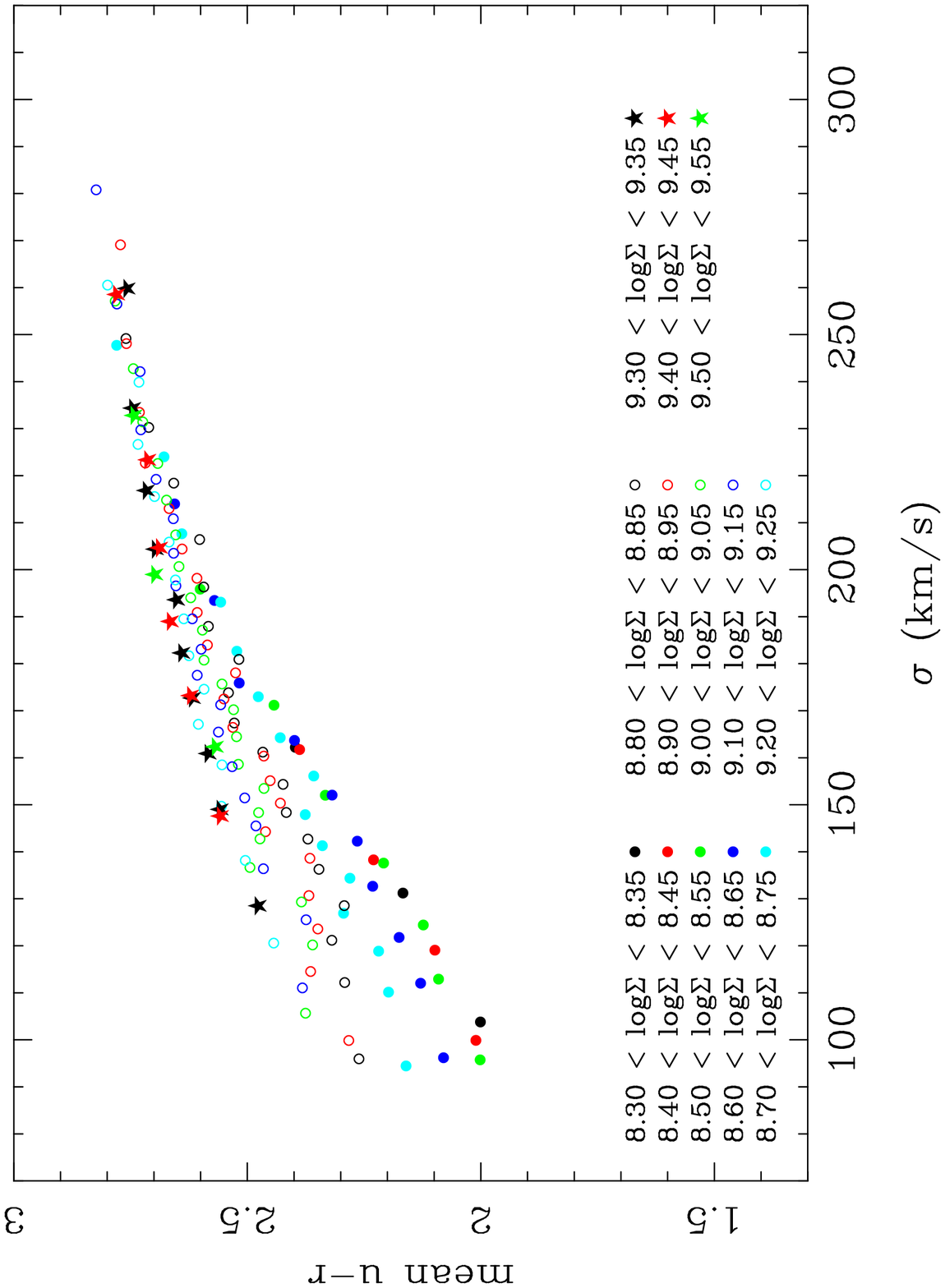}
\includegraphics{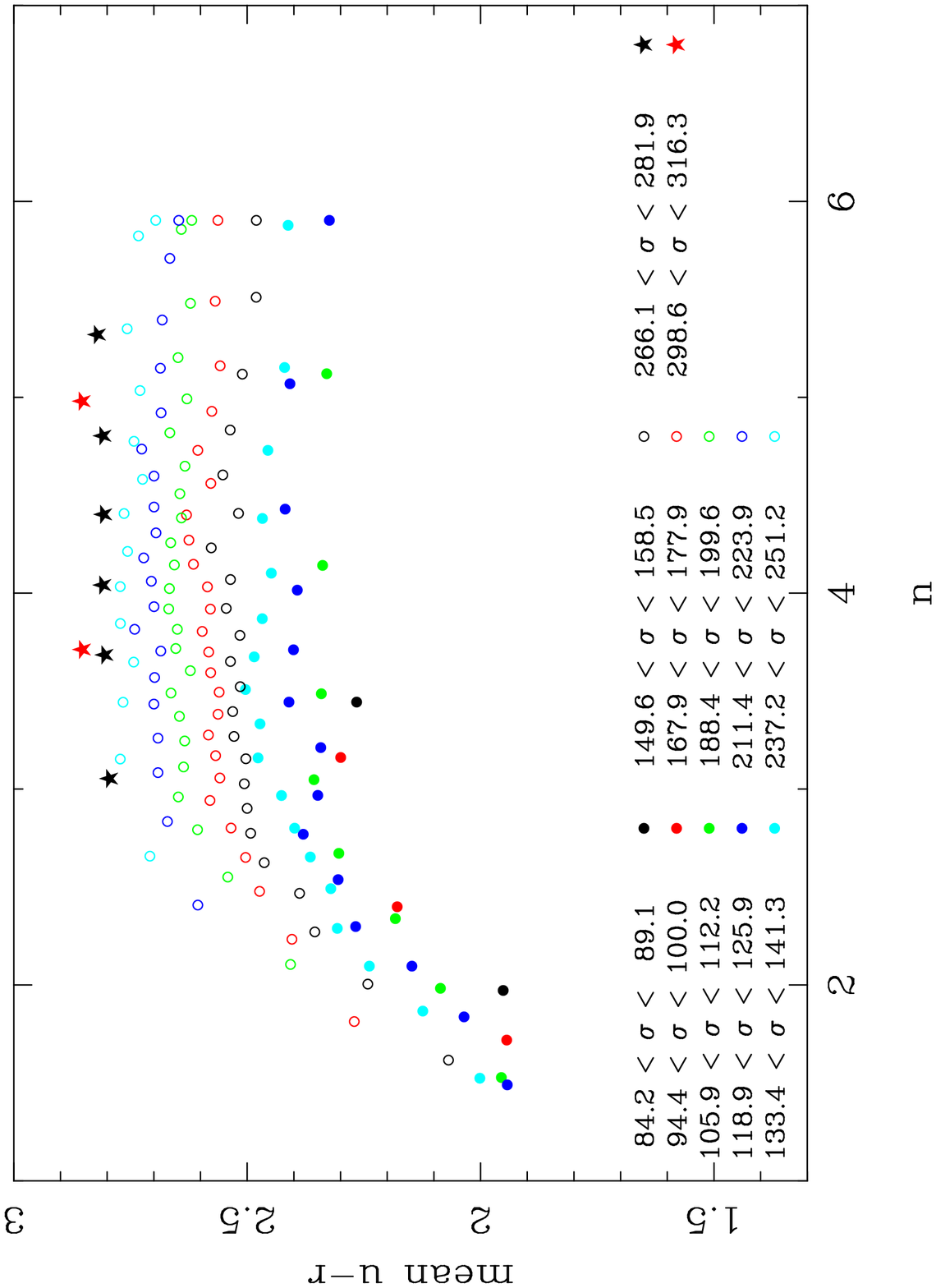}
\includegraphics{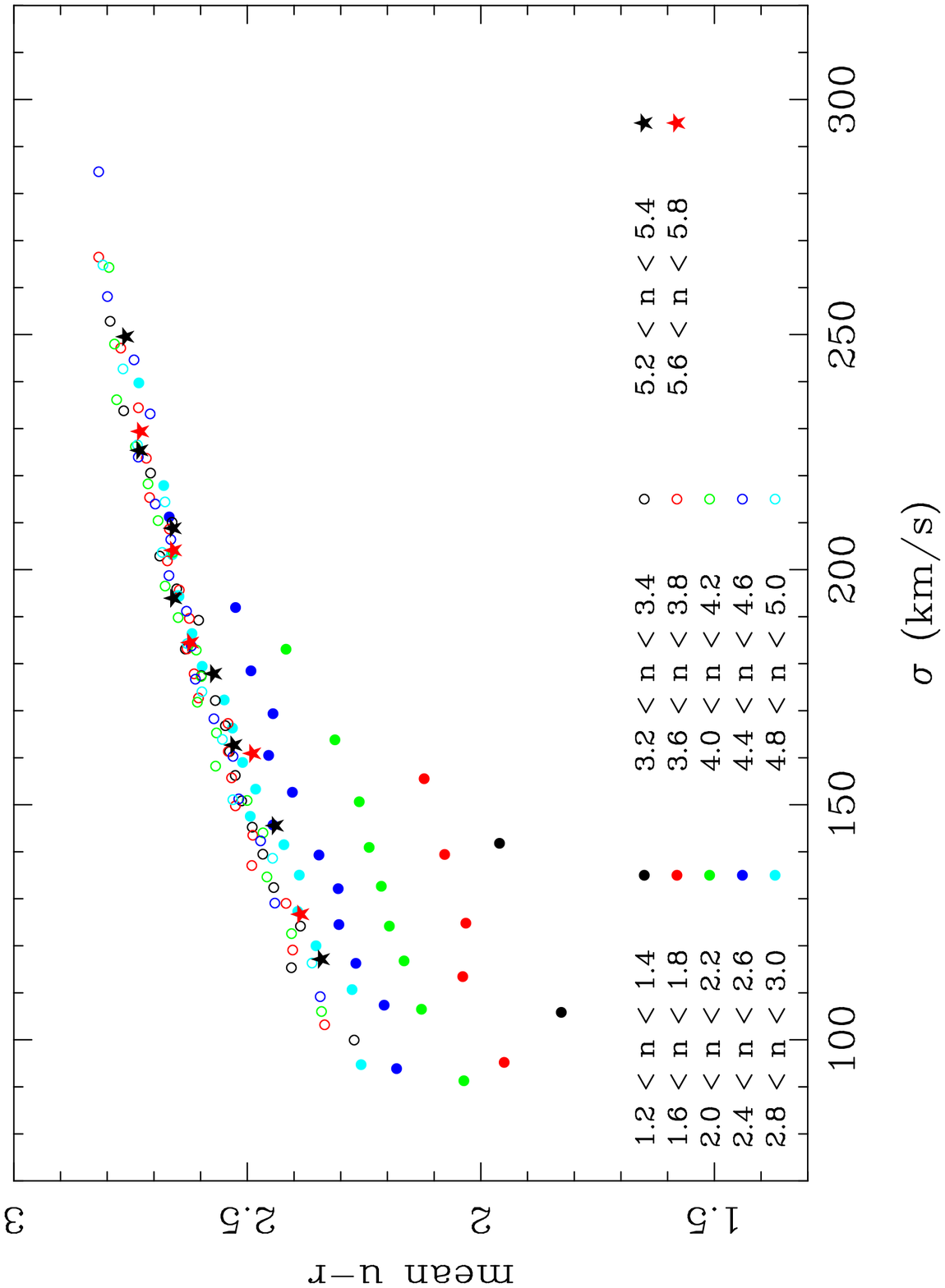}
\caption{\small The relationship between mean $u-r$ color and velocity dispersion at fixed stellar mass (top left), stellar mass at fixed velocity dispersion (top right), velocity dispersion at fixed mass surface density (middle left) and mass surface density at fixed velocity dispersion (middle right), velocity dispersion at fixed sersic n (bottom left) and sersic at fixed velocity dispersion (bottom right). 
\label{fig:colVdispNsMdenSM}}
\end{figure*}

We wish to understand how the typical color of a galaxy depends on its
stellar mass, central velocity dispersion,
surface mass density, and Sersic profile. We begin in Figure
\ref{fig:ColNsVdMdSM} were we show the $u-r$ color distribution for
all SDSS galaxies with 0.02 $< z <$ 0.11 as a function of each
parameter. We have applied to this figure (and all subsequent figures)
a V/Vmax weight for each galaxy to correct for the varying stellar
mass completeness limit with redshift. We plot the color distributions
as a function of log(\Mstar), log(\Mden), log(\Vdisp$^2$) and
log($n^2$).

We have chosen these units as they are all approximately
linearly proportional to one an other.
This is illustrated in Figure
\ref{fig:VdispMdenSM} were we show the relationships between these four
parameters. The contours show the full distributions and the points show the mean of $y$ in bins of $x$ (black) and $y$ in bins of $x$ (red).
Whilst we may expect \Mden\ and \Vdisp$^2$ to be approximately linearly proportional
to \Mstar\ and each other, we find that $n^2$ shows the same trends over a broad range in those parameters. Therefore we can meaningfully compare the
predictive power of $n^2$ to that of the other three parameters.

Figure \ref{fig:ColNsVdMdSM} shows that there is a clear color
dependence on each parameter but with significant scatter. It also
illustrates, as has been previously found (e.g., Kauffmann et al.\
2003), that \Mstar\ is a relatively poor
discriminator of a galaxy's color, particularly around the
``transition mass'' of $\sim 3\times 10^{10}\,\Msun$.
It is important to note that
distributions become noisy at both low \Mstar~and low
\Vdisp. The number of galaxies becomes increasing small at low masses,
due to the apparent magnitude limit of the SDSS, which also strongly
affects the sample at low \Vdisp\ due to the tight correlation between
the two parameters. \Vdisp\ measurements also become more uncertain
below 65 km/s due to the resolution of the SDSS spectra. For these
reasons we limit the remainder of the analysis to galaxies with
\Mstar~$> 10^{10} \Msun$ and \Vdisp~$>$ 65 km/s. 

We also remove edge on disk galaxies from our sample to minimize the influence of dust on our measurements \citep[e.g.,][]{Patel11}, although we expect the dust contribution to be low in the local universe. We make a cut using the Galaxy Zoo $P_{edge}$ parameter limiting it to be $<$ 0.3. We note that this cut changes our results very little and has no effect on any of our conclusions. 
  
\section{Which parameter correlates best with color?}

In order to determine which parameter shows the strongest correlation
with color
we determine whether there remains any color dependence on \Mstar,
\Mden, \Vdisp, and $n$ when each of these parameters are held
fixed. We divide our parent sample ( 0.02 $< z <$ 0.11, \Mstar~$>10^{10}\Msun$, \Vdisp\ $>$ 65 km/s, $P_{edge} <$ 0.3) into a
series of samples selected to have narrow ranges in each parameter. We
select bins of 0.05 in log(\Mstar) and log(\Mden), 0.025 in
log(\Vdisp) and 0.2 in $n$. We then calculate the mean $u-r$ color of
galaxies in each of these narrow binned samples as of function of the
other three parameters, where the mean is calculated in bins of 300
galaxies.

\subsection{Residual Correlations}
 
We show the resulting relationships between mean color and \Mstar,
\Mden, $n$ and \Vdisp~in Figures \ref{fig:colNsMdenSM} and
\ref{fig:colVdispNsMdenSM}.  Each row shows a pair of parameters with
the binning and abscissa parameter switched. That is, in each
row the left
panel shows the effect of varying parameter 1 at fixed parameter 2,
and the right panel shows the effect of varying parameter 2 at fixed
parameter 1.
The left and right panel
therefore essentially show the same information, but highly the trends in a complimentary way.
If the trends of all of the
binned samples lie on top of each other in any of the
plots it would mean that the color
would be independent of the parameter used for the binning. Similarly,
if the abscissa parameter is unimportant each
of the individual binned trends will be flat.

Turning to the first pair of parameters at the top of Figure
\ref{fig:colNsMdenSM}, \Mden~and \Mstar, we can see something close to
this extreme situation. When \Mstar~is held fixed (left panel) there
remains a clear trend with \Mden, such that higher \Mden~galaxies are
redder. The trend is particularly strong at low \Mstar~and gradually
decreases at the highest masses, resulting in some spread but a
generally low dispersion between the \Mstar~bins. The same trend is
visible when \Mden~is held fixed; there is only a very weak dependence
of mean $u-r$ color on \Mstar~with more massive galaxies being
redder. The individual \Mden~bins are well separated, showing the
strong color dependence on \Mden, and are generally parallel. It
appears that \Mden~is a better indicator of a galaxy's color than its
stellar mass is.

We next consider \Mstar~and $n$, shown in the middle panel of Figure
\ref{fig:colNsMdenSM}. For low $n$ ($<$ 2.5) there is a strong
dependence of the color on $n$ at fixed \Mstar, but for higher $n$ the
trends with \Mstar~are all the same with no $n$ dependence. At all $n$
there is a trend for more massive galaxies to be redder on average,
although this is very weak for galaxies with log(\Mstar) $<$ 11. Again
\Mstar~appears a poorer indicator of a galaxies color than $n$,
although this is less true at high $n$ or high \Mstar.

The bottom panels of Figure \ref{fig:colNsMdenSM} concern \Mden~and
$n$. At low $n$ there is a very strong relationship between color and
\Mden~with higher \Mden~galaxies being redder. Whilst this trend remains
for all $n$ the slope of this correlation reduces as $n$
increases. There is a similar dependence of color on $n$ at fixed
\Mden, although the relation becomes very weak at the highest \Mden.

In Figure \ref{fig:colVdispNsMdenSM} we show how the color depends on
\Vdisp~at fixed \Mstar, \Mden, and $n$ on the left and how color
depends on \Mstar, \Mden, and $n$ at fixed \Vdisp~on the right. It is
striking how tight all of the trends are in the left panels and how
separated they are in the right panels. This indicates that the mean
galaxy color is more strongly dependent on \Vdisp~than \Mstar,
\Mden~or~$n$. This is particularly true at \Vdisp~($>$ 200 km/s) where
all the trends lie almost completely on top of each other on the left
and are very close to flat on the right. At lower \Vdisp\
interesting trends
emerge; perhaps surprisingly the mean color becomes bluer as
\Mstar~increases, with this trend increasing as \Vdisp~decreases. This
probably reflects an increasing disk component in more massive galaxies
at fixed \Vdisp.
At low \Vdisp\ ($<$175 km/s) a trend emerges with \Mden~such that
galaxies with higher \Mden~at fixed \Vdisp~have redder colors, with
the trend becoming more significant as \Vdisp~decreases. There is also
a trend with $n$ when $n <$ 2.5 such that galaxies with higher $n$ are
redder. There is no color dependence on $n$ at fixed \Vdisp~for higher
$n$ galaxies. Whilst some trends with \Mstar, \Mden~and $n$ emerge in
some regions of the parameter space there is always a strong
dependence of the mean color on \Vdisp~over the full range of the
other parameters.

\subsection{Quantifying the Residual Correlations}

\begin{figure*}
\vspace{14.0cm}
\includegraphics{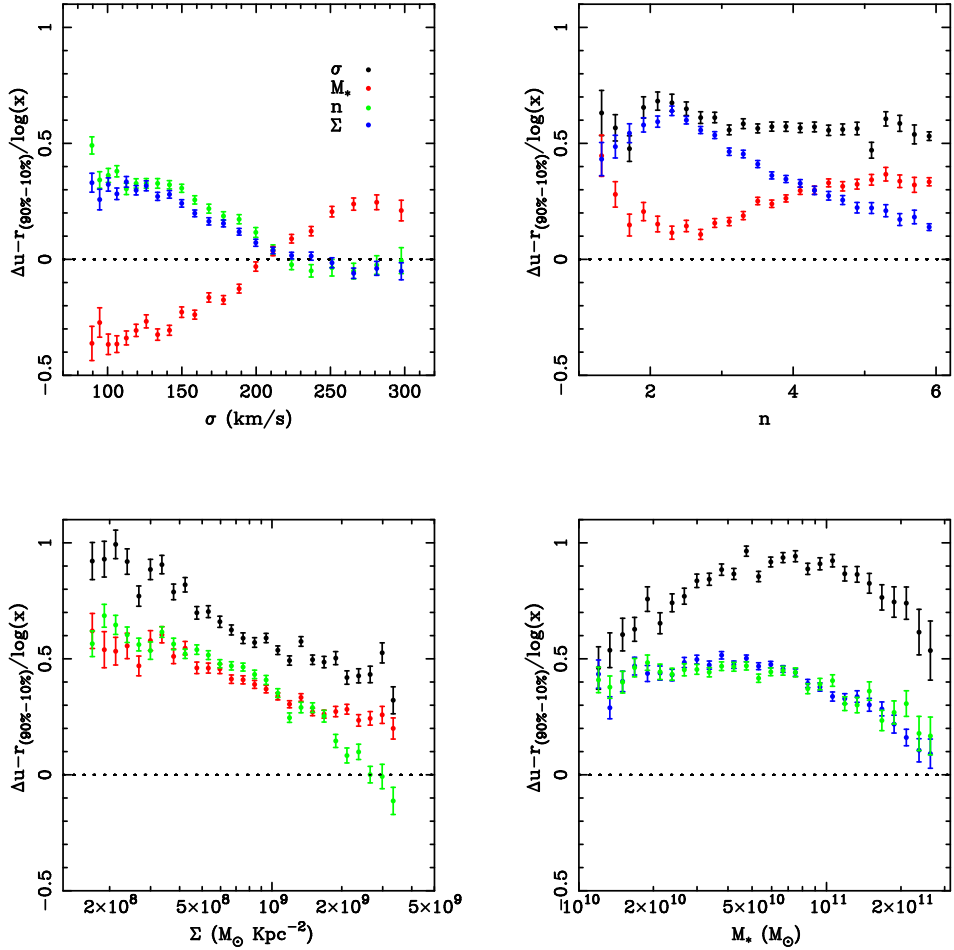}
\caption{\small The gradient in the dependence of the mean $u-r$ color on each parameter (\Mstar, \Mden, \Vdisp and $n$) when each is fixed in turn. This is calculated as the difference in $u-r$ color between the 90th and
  10th percentile of parameter x per unit parameter x when a second parameter is fixed. The units of parameter x are log(\Mstar), log(\Vdisp$^2$),
  log(\Mden) and log($n^2$) which are all roughly equivalent. A color
  difference of zero means that there is no change in the color as
  parameter x is varied when the second parameter is fixed, as is the case for
  $n$ and \Mden~when \Vdisp~is fixed at values $>$220 km/s (top
  left). Positive values indicate that galaxies become redder as
  parameter x is increased. When \Vdisp~is
  fixed (top left) the absolute color variation over the other
  parameters is always less than 0.5. When the other parameters are
  fixed the magnitude of the color variation with \Vdisp~is always
  largest and is almost always larger than 0.5. The implication is
that color depends more strongly on \Vdisp~than on any other
parameter.
\label{fig:dcol}}
\end{figure*}

We show in Figure \ref{fig:dcol} an attempt to both simplify and
quantify the trends that are displayed in Figures
\ref{fig:colNsMdenSM} and \ref{fig:colVdispNsMdenSM}. In narrow bins
of one parameter we calculate the difference in the mean $u-r$ color
of the galaxies lying in the lowest and highest 10 percentiles of the
other three parameters. Since the size of the range of the second
parameter may vary with the first and between the different parameters
we calculate this color difference per unit log(\Mstar), log(\Mden),
log(\Vdisp$^2$) and log($n^2$). For example the blue points in the top
left panel of \ref{fig:dcol} show the difference between the mean
$u-r$ color of galaxies with the 10\% highest \Mden~and 10\% lowest
\Mden~per unit log(\Mden) in narrow bins of \Vdisp. This is
essentially the vertical scatter in the center left panel of Figure
\ref{fig:colVdispNsMdenSM} or the gradient of the individual trends in
the center right panel.

The color gradients shown in Figure \ref{fig:dcol} reinforce the
conclusions we have already drawn. When \Vdisp~is fixed (top left
panel) the magnitude of the color gradient is always less than 0.5,
showing that there is only a weak dependence of color on \Mstar,
\Mden~or $n$ at constant \Vdisp. This is particularly true at high
\Vdisp~($>$ 200 km/s) where the gradient is essentially zero for both
\Mden~and $n$. When the other three parameters are held fixed the
color gradient with \Vdisp~is always the largest and is greater than
0.5 regardless of the values of the other parameters. Clearly there is
a significantly larger dependence of the color of galaxies on
\Vdisp~than on \Mstar, \Mden~or $n$.

Figure \ref{fig:dcol} again confirms the negative color gradients with
\Mstar~at low \Vdisp~(red points in top left panel), such that more
massive galaxies are bluer when \Vdisp~is fixed. We also see
negative color gradients with $n$ for the highest
\Mden~galaxies. Otherwise it is always the case that galaxies with
higher \Mstar~or $n$ are redder, and always the case that galaxies
with higher \Vdisp~or \Mden~are redder when the other parameters are
fixed.

\section{Discussion}

The central result of this paper is that the colors of galaxies
depend more strongly on
\Vdisp\ than on \Mstar, \Mden, or $n$. We have demonstrated this by
examining, for each of these parameters, how strong residual correlations
with the other three parameters are when the parameter under consideration
is held fixed. At fixed \Vdisp\ residual trends with other parameters
are weak, and when any of the other parameters is fixed there are strong
residual trends with \Vdisp.

Recently \citet{Bell11} studied how the fraction of passive (i.e., not
star-forming) galaxies depends on a similar set of parameters and
reached the conclusion that $n$ was the best indicator of the passive fraction. This seems to
be at odds with our findings where \Vdisp~shows the strongest color
trends. The difference may stem from the fact that
\citet{Bell11} use \Mstar/\Re\ to approximate
velocity dispersion. Whilst there is a correlation between \Vdisp\ and
\Mstar/\Re, which is improved if corrections based on $n$ are applied
\citep{Taylor10}, there is
significant scatter ($\sim$0.1 dex). This may be
the cause of our differing conclusions and
illustrate the importance of directly measured dynamical properties.

Our results, and those of others \citep[e.g.,][]{Kauffmann03,
Franx08, Dokkum11, Bell11},
call into question whether ``mass quenching'' \citep[e.g.,][]{Peng10} and other mass-driven effects are actually manifestations of
underlying trends with velocity dispersion. The velocity dispersion
may in turn reflect a yet more fundamental parameter. It is
known to correlate well with central black hole mass \citep[e.g.,][]{Ferrarese00, Gebhardt00} and also with the properties of dark matter
halos \citep{Wake12}. Taken together, all these results suggest
a simple scenario where the mass of the dark matter halo determines both
the central black hole mass and the star formation history of galaxies.

This study can be extended and improved in many ways. The dispersions
are currently corrected to a common aperture with reasonable
assumptions, but it would be very
useful to measure radial trends in dispersion in a systematic way.
This is particularly relevant for low mass galaxies and star forming
galaxies, as they have significant disks which presumably dominate
at large radii. Blue disks may well be the cause of the peculiar
fact that, at low \Vdisp, more massive galaxies are {\em bluer} at
fixed dispersion (see Fig.\ 4). Modeling of the effects of various
physical processes (such as merging) on the velocity dispersion may
help us understand {\em why} velocity dispersion is so well correlated
with many aspects of galaxies. One possibility is that \Vdisp\ may well be indicating both the halo mass (or other halo properties), in a similar or more precise manner than \Mstar\ \citep{Wake12}, and at the same time be an indicator of the relative bulge to disk components. So at fixed \Mstar\ a higher \Vdisp\ galaxy has a larger bulge to disk ratio and so is redder, whereas at fixed $n$ a higher \Vdisp\ galaxy typically occupies a more massive dark matter halo and so is also redder. \Vdisp\ is then the best of the four parameters at encapsulating both color dependences (halo mass and bulge to disk ratio) as illustrated by the tightness of the \Vdisp\ - \Mstar\ and \Vdisp\ - $n$ relations shown in Figure \ref{fig:VdispMdenSM}.
 
Finally, 
it would be very interesting
to do similar systematic studies at earlier cosmic epochs, extending those pioneered by \citet{Franx08} and
\citet{Bezanson11}, which will give further insight into the physical parameters which
drive galaxy formation and evolution.

\end{document}